\documentclass[aps,prb,twocolumn,showpacs,a4paper,floatfix]{revtex4}

\usepackage{amsfonts}
\usepackage{amsmath}
\usepackage{dcolumn}
\usepackage{amssymb}
\usepackage{mathtools}
\usepackage{amsbsy}
\usepackage{textcomp}
\usepackage[english]{babel}
\usepackage{graphicx}
\usepackage{pgf}
\usepackage{tikz}
\usepackage{verbatim}

\begin{document}
\title{Interplay between bending and stretching in carbon nanoribbons}
\author{Emiliano Cadelano, Stefano Giordano, Luciano Colombo}
\email[e-mail to: ]{luciano.colombo@dsf.unica.it}
\affiliation{Dipartimento di Fisica, Universit\`a di Cagliari\\ Cittadella Universitaria, I-09042 Monserrato (Ca), Italy}
\date{\today}

\begin{abstract}
{ 
We investigate the bending properties of carbon nanoribbons by combining continuum elasticity theory and tight-binding atomistic simulations. First, we develop a complete analysis of a given bended configuration through continuum mechanics. Then, we provide by tight-binding calculations the value of the bending rigidity in good agreement with recent literature. We discuss the emergence of a stretching field induced by the full atomic-scale relaxation of the nanoribbon architecture. We further prove that such an in-plane strain field can be decomposed in a first contribution due to the actual bending of the sheet and a second one due to the edges effects induced by the finite size of the nanoribbon.  
}
\pacs{62.25.-g, 62.20.D-, 46.70.Hg}
\end{abstract}
\maketitle

\section{Introduction}

Graphene\cite{novoselov} plays a unique role in materials science since it is the mother structure of most carbon $sp^{2}$ nanosystems of current interest. By stacking, folding or bending a graphene sheet it is indeed possible to generate, respectively, graphite-like  systems, fullerene cages (pentagonal rings are here needed as well) or nanotubes.
In particular, the bending properties are critical in attaining the structural stability and morphology for both suspended and supported graphene sheets, and directly affect their electronic properties.\cite{RMP} Rippling of pure graphene has been also observed with mesoscopic amplitude and wavelength, both for suspended monolayers\cite{meyer} and sheets deposited on substrates such as silicon dioxide.\cite{ishi} 
Moreover, the bending properties play a central role in the design of graphene- or carbon nanotube-based devices, like e.g. mechanical resonators.\cite{res1,res2} The bending features of functionalized graphene sheets have been probed by atomic force microscopy, observing that the folding behavior is dominated by defects and functional groups.\cite{schniepp} 
Finally, bending ultimately governs the carbon nanotubes unzipping process, recently used to produce narrow ribbons for nanoelectronics.\cite{kosy} With the same technique, a new class of carbon-based nanostructures, which combine nanoribbons and nanotubes, has been introduced in order to obtain magnetoresistive devices.\cite{santos}

Within this scenario we frame the present investigation, addressed to improve our fundamental understanding of the bending properties of a one-atom thick carbon sheet. 
The main goal is twofold: (i) to draw a thorough theoretical picture on bending, fully exploiting the elasticity theory and providing an atomistic quantitative estimation of the corresponding bending rigidity; (ii) to prove that the bending process of a carbon nanoribbon is always associated with the emergence of a (small) stretching, particularly close to the edges. These results have been obtained by combining continuum elasticity theory and tight-binding atomistic simulations (TB-AS). 

The conceptual development and actual exploitation of our theoretical model proceeds through the following steps. 
At first, by means of continuum mechanics we have obtained the exact shape for a purely bended nanoribbon, by imposing suitable boundary conditions.
The bending rigidity is then evaluated by TB-AS for several nanoribbons differing by length and width.
As a second step, we observed that, under the above assumption of pure bending, the corresponding rigidity must be a constant independent of the actual shape of the sheet. Nevertheless by allowing full atomic-scale relaxation during bending, we rather found a geometry-dependent rigidity, a feature that we have attributed to the onset of stretching phenomena.
Therefore, as final step, we have developed a procedure to discriminate between stretching and bending energy, so providing a complete picture about the mechanical behavior of graphene and also reconciling the atomistic data with the continuum theory results. 

The structure of the paper follows: in Section II we outline the theoretical framework from both the continuum elasticity theory and the tight-binding atomistic simulations point of view. In Section III we describe the results concerning the bending stiffness and the interplay between stretching and bending. Finally, in Section IV we draw the conclusions.

\section{Theoretical framework}

\subsection{Continuum picture}
The graphene strain energy density $\mathcal{U}$  [eV\AA $^{-2}$] is defined as \cite{green, landau} 
\begin{eqnarray}
\label{eq:density energy}
\nonumber
\mathcal{U}&=& \frac{1}{2}\frac{E}{1+\nu} \mbox{Tr}\left( \hat{\varepsilon}^{2}\right) +\frac{1}{2}\frac{E\nu}{1-\nu^{2}}  \left[ \mbox{Tr}\left( \hat{\varepsilon}\right)\right] ^{2} \\
&&+\frac{1}{2}\kappa\left( 2\mathcal{H}\right)^{2}-\bar{\kappa}\mathcal{K} 
\end{eqnarray}
where $E$ [Nm$^{-1}$] and $\nu$ are the two dimensional Young modulus and the Poisson ratio, while $\kappa$ [eV] and $\bar{\kappa}$ [eV] are the bending rigidity and the Gaussian rigidity, respectively. The in-plane deformation (stretching) energy [given by the first two terms in Eq.(\ref{eq:density energy})] is described by the standard small strain tensor $\hat{\epsilon}=\frac{1}{2}(\vec{\nabla}\vec{u}+\vec{\nabla}\vec{u}^{\rm T})$, being $\vec{u}$ the displacement field. On the other hand, the out of plane deformation (bending) energy [given by the last two terms in Eq.(\ref{eq:density energy})] is described by the mean curvature $\mathcal{H}=\tfrac{k_{1}+k_{2}}{2}$ [m$^{-1}$] and by the Gaussian curvature $\mathcal{K}=~k_{1}k_{2}$ [m$^{-2}$], where $k_{1}$ and $k_{2}$ are the principal curvatures at a given point on the surface.\cite{do carmo} They are straightforwardly given by $k_{1}=1/R_1$ and $k_{2}=1/R_2$ where $R_1$ and $R_2$ are the principal radii of curvature at that point.
In the case of a continuum plate of thickness $h$ made of an isotropic and homogeneous material, the classical Kirchhoff theory provides $\kappa=\tfrac{1}{12}\frac{Eh^{2}}{1-\nu^{2}}$ and $\bar{\kappa}=\tfrac{1}{12}\frac{Eh^{2}}{1+\nu}$ (note that $ E=Yh $ where $ Y $ is the three-dimensional Young modulus).\cite{landau}
For an infinitesimally thin graphene monolayer such a theory does not apply, since the thickness $h$ cannot be unambiguously defined and the bending moment has simply a different physical origin. While the bending moment for the Kirchhoff plate derives from a compression/extension of the different material layers forming the thickness $h$, in graphene it is due to the interactions among orbitals $p_{z}$ which are affected by the bending process.
Therefore, the determination of $\kappa$ and $\bar{\kappa}$ for graphene is a well-posed (and, to a large extent, still open) problem, which is independent of the evaluation of $E$ and $\nu$.\cite{arroyo3}

\begin{figure}[t]
\label{fig:geometri3D}
\centering\includegraphics[width=0.35\textwidth]{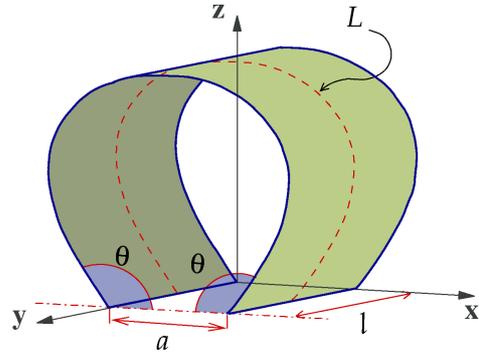}
\caption{\label{capt:geometri3D}(Color online) Bended ribbon with length $l$ and width $L$ (red dashed line). The parallel edges with length $l$ are fixed at distance $a$, while the attack angles $\theta$ is free to relax.}
\end{figure}

Our model system is a rectangular ribbon with length $ l $ and width $L$ (see Fig.\ref{capt:geometri3D}). The ribbon is bended without stretching ($ \hat{\varepsilon}=0 $) along its width. The boundary conditions consist in fixing the positions of the two parallel edges (with length $ l $) at a given distance $a$, while the attack angles $\theta$ is free to relax. This configuration involves only one curvature $k_{1}$, leading to $\mathcal{H}=\tfrac{k_{1}}{2}$ and $\mathcal{K}=0$. 
By considering different values of $ a $ in the range $ (0,L) $ we obtained a set of differently bended configurations. The elastic problem consists in finding the sheet shape by minimizing the bending energy 
\begin{equation}
\label{eq:energy integral}
 {U}_{b}=\int\int_{A}\mathcal{U}dA=\frac{1}{2}\kappa l\int_{0}^{L}k_{1}^{2}ds
\end{equation}
where $ A=Ll $ is the total area of the system. If the configuration is described by the function $ z=z(x) $, then we have $ k_1=z''/\left[1+ \left(z' \right)^{2} \right] ^{3/2} $, where $ z'=dz/dx $ and $z''=d^2z/dx^2$. On the other hand, $ ds=\sqrt{g}dx $ where $ \sqrt{g}=\sqrt{1+ \left(z' \right)^{2} } $. Therefore, Eq.(\ref{eq:energy integral}) assumes the explicit form 
\begin{equation}
\label{eq:energy integral2}
 {U}_{b}=\frac{1}{2}\kappa l\int_{0}^{a}\frac{\left( z''\right)^{2} }{\left[1+ \left(z' \right)^{2} \right] ^{5/2}} dx
\end{equation}
The problem consists in finding the curve $ z=z(x) $ minimizing the energy functional in Eq.(\ref{eq:energy integral2}) under the constraint  
\begin{equation}
\label{eq:energy integral3}
 \int_{0}^{a}\sqrt{1+ \left(z' \right)^{2} } dx=L
\end{equation}
enforcing the absence of any in-plane stretching.

By the application of the constrained variational calculus we eventually obtain the final geometry in parametric representation $ [x(s),z(s)] $
\begin{eqnarray}
\label{eq:parametric geometry}
\frac{x}{L}&=& \frac{\mathsf{E}(q) -\mathcal{E}\left(\mathrm{am}\left\lbrace \mathsf{K}(q)\left( 1-2\dfrac{s}{L}\right)\right\rbrace,q \right)}{\mathsf{K}(q)} -\frac{s}{L}
\\
\label{eq:parametric geometry2}
\frac{z}{L}&=& \frac{q}{\mathsf{K}(q)}\mathrm{cn}\left\lbrace \mathsf{K}(q)\left( 1-2\dfrac{s}{L}\right)  \right\rbrace 
\end{eqnarray}
where $s$ is the arc length ($0<s<L$), $q=\sin{\tfrac{\theta}{2}}$ is the elliptic modulus and $\theta$ is the attack angle given by
\begin{eqnarray}
\label{eq:edge distance}
\frac{a}{L}=2\frac{\mathsf{E}(q)}{\mathsf{K}(q)} -1.
\end{eqnarray}
The quantities $\mathsf{E}(q)$ and  $\mathsf{K}(q)$ are the complete elliptic integrals, defined as\cite{grad,abra}
\begin{eqnarray}
\mathsf{E}(q)={\cal{F}}\left(\frac{\pi}{2},q \right), \,\,\,\,\,\,\, \mathsf{K}(q)={\cal{E}}\left(\frac{\pi}{2},q \right)
\end{eqnarray}
where the functions $ {\cal{F}}(v,q)$ and ${\cal{E}}(v,q) $ are incomplete elliptic integrals of the first and second kind, respectively \cite{grad,abra}
\begin{eqnarray}
\nonumber
 {\cal{F}}(v,q)&=&\int\limits_0^v {\frac{d\alpha }{\sqrt {1-q^2\sin ^2\alpha } }}  \\ 
 {\cal{E}}(v,q)&=&\int\limits_0^v {\sqrt {1-q^2 \sin ^2\alpha } d\alpha }.
\end{eqnarray}
Moreover, by considering $ u={\cal{F}}(v,q) $ we define the inverse relation (with fixed modulus $ q $) $v=\mathrm{am}\left\lbrace u \right\rbrace$, which is called Jacobi amplitude function. Further, $\mathrm{cn}\left\lbrace u \right\rbrace=\cos v= \cos\left( \mathrm{am}\left\lbrace u \right\rbrace\right) $  and $\mathrm{sn}\left\lbrace u \right\rbrace=\sin v= \sin\left( \mathrm{am}\left\lbrace u \right\rbrace\right) $ are the Jacobi elliptic functions.\cite{abra}
Interesting enough, one can prove that $\lim_{a/L \rightarrow 0}\theta=~130.709^{o}$, an universal value of the attack angle found whenever $a=0$ or $L$ is very large.

\begin{figure}[t]
\label{fig:72}
\centering\includegraphics[width=0.48\textwidth]{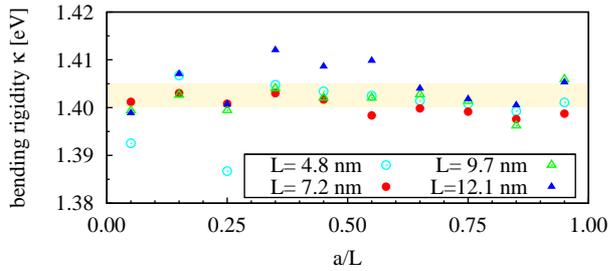}
\caption{\label{capt:72}
(Color online) Bending rigidity $\kappa$ obtained for purely ribbons with several widths $L$. The average value is given by $\kappa_{ave}=~1.4025\pm0.0025$ eV (yellow area shows the error bar)}
\end{figure}

\subsection{Atomistic simulations}

The present TB-AS \cite{colombo} have been performed making use of the sp$^3$, orthogonal, and next-neighbors tight-binding representation by Xu et al.\cite{xu} 
The present TB total energy model has been implemented within the scheme given by Goodwin \textit{et al.}\cite{good} for the dependence of the TB hopping integrals and the pairwise potential on the interatomic separation.

Applications to molecular-dynamics studies of liquid carbon
and small carbon clusters indicate that this model correctly describes carbon systems over a wide range of environments.\cite{xu}
This approach has been successfully used to show that the surface of nanodiamond particles reconstructs in a fullerenelike manner, generating carbon clusters called bucky diamonds.\cite{galli}
Moreover, the growth of nanostructures (linear, ring, and fullerenelike objects) in a carbon plasma\cite{Yamaguchi} and 
the formation of carbon clusters (onion-like and endohedral structures) from the condensation of liquid droplets\cite{Bogana} have been simulated by the present tight-binding model.
Finally, this parametrization has been recently used for determining the nonlinear elastic moduli governing the graphene stretching elasticity.\cite{cadelano}

The previous continuum analysis is useful both to create the input configurations for atomistic calculations and to define the simulation protocol. 
The investigated system consists in a nanoribbon formed by a perfect hexagonal carbon lattice, having width $L$ in the range 4-12 nm and length $l$  imposed to obtain a simulation box containing a constant number of $\sim 600$ carbon atoms. Morever, periodic boundary conditions are assumed along the direction of the length $l$.
The length (width) is developed along the armchair (zig-zag) direction of the honeycomb lattice.
Each nanoribbon is deformed as defined in Eqs.(\ref{eq:parametric geometry}) and (\ref{eq:parametric geometry2}) in ten configurations corresponding to different values of $a$. In any bended configuration, all the interatomic distances are fixed at the equilibrium value for flat graphene (so that no bond stretching is for the moment allowed).
The bending rigidity has been straightforwardly obtained as $\kappa=\tfrac{2}{l \mathcal{I}}{U}_{b}$ with ${U}_{b}$ given by Eq.(\ref{eq:energy integral}), where the integral $\mathcal{I}=\int_{0}^{L}k_{1}^{2}ds$ is computed for the given configuration. It is important to remark that the obtained value for $\kappa$ must be independent of the actual configuration since the deformation is a pure bending one.

\section{Results}

Accordingly to the scheme outlined in the previous Section, we have firstly evaluated the (pure) bending energy as $  {U}_{b}= E_{o}^{bended} -E_{o}^{flat}$, where $ E_{o}^{bended} $ and $E_{o}^{flat}$ represent the TB-AS total energy of the bended (but not relaxed) and equilibrium (flat) configurations, respectively.
The atomistic results for $\kappa$ are reported in Fig.\ref{capt:72} (symbols) as function of the $a/L$ ratio and for different width $L$. We estimate an average value $ \kappa_{ave}=1.40$ eV. While the reported values of $\kappa$ (for nanotubes) vary in the range 1 eV$\lesssim \kappa \lesssim$2 eV, \cite{tu} we remark the most reliable \textit{ab-initio} data $\kappa= ~1.40$ eV, \cite{arroyo3} and $ \kappa=1.46$ eV, \cite{kudin} are in excellent agreement with our prediction, a feature standing for the reliability of the present computational procedure.

\begin{figure}[t]
\label{fig:graphene angles1}
\centering\includegraphics[width=0.47\textwidth]{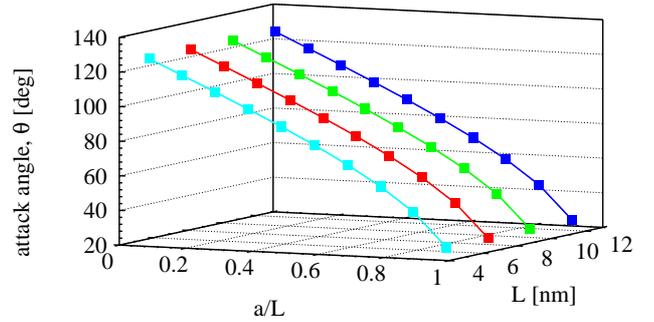}
\caption{\label{fig:graphene angles} (Color online) The theoretical results for the values of the attack angles $\theta$ (circles), predicted from Eq.(\ref{eq:edge distance}) for several ribbon with different width $L$, each at different edge-distances $a$, are compared with the corresponding data from atomistic simulations (crosses).}
\end{figure}

Although reassuring, the above picture must be refined in order to properly take into account atomic-scale features. Therefore, full relaxation of the internal degrees of freedom of the bended systems is performed by zero temperature damped dynamics until interatomic forces resulted not larger than $\sim 10^{-5}$eV/\AA. We have so generated a new set of configurations where bending and stretching features are entangled. 
During the relaxation, the positions of the atoms belonging to the edges (i.e. atoms with $x=0$ or $x=a$, see Fig.\ref{capt:geometri3D}) are fixed and, therefore, the distance  $a$ between the edges remained constant.
Overall we observed that the geometry is only marginally affected by relaxation as shown in Fig.\ref{fig:graphene angles}. Here we compare the attack angle $\theta$ predicted from Eq.(\ref{eq:edge distance}) versus the ratio $a/L$ with the corresponding values obtained from the relaxed configurations.  We note that, for $ a/L\rightarrow 0 $, we obtain the universal value $130.709^{o}$ as previously discussed. As a matter of fact, after the relaxation, the attack angle $\theta$ do not change and the maximum variation of $L$ was as little as  $0.005$ nm, corresponding to a variation of the integral $\mathcal{I}$ smaller than $0.01\%$.
Nevertheless, even for such minor relaxations the energetics of the fully relaxed systems is expected to sizeably differ from the purely bended case, because of the extraordinary large value of the graphene Young modulus.\cite{cadelano} It is therefore important to provide a new estimation of the bending energy for the fully relaxed configurations.

\begin{figure}[t]
\label{fig:bending rigidity1}
\includegraphics[width=.49\textwidth]{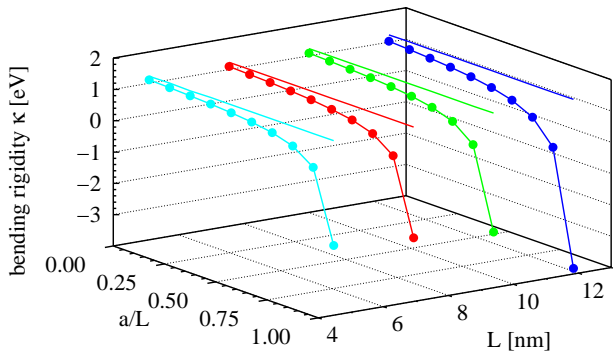}
\caption{\label{fig:bending rigidity}(Color online) Bending rigidity $\kappa$ computed by means of ${U}_{b}= E_{relaxed}^{bended}-E_{relaxed}^{flat}$ (full circles). Straight lines correspond to the average value $\kappa_{ave}= ~1.40$ eV as deduced from Fig.1.}
\end{figure}

Following the above argument, we evaluated the new bending rigidity $\kappa$ by means of the energy ${U}_{b}= E_{relaxed}^{bended}-E_{relaxed}^{flat}$ and Eq.(\ref{eq:energy integral}), where $E_{relaxed}^{bended}$ is the energy of a relaxed bended ribbon and $E_{relaxed}^{flat}$ is the energy of a flat ribbon after a full relaxation (different from the energy of the infinite graphene sheet because of the edge effects).
In this case, we have found a variation of $\kappa$ upon $a/L$ as shown in Fig.\ref{fig:bending rigidity} (full circles).
This result suggests that atomic-scale relaxations upon bending have induced as expected an additional field of in-plane stretching, which provides new energy contributions as reported in Eq.(\ref{eq:density energy}).
It is interesting to observe that the largest differences between the unrelaxed and relaxed configurations are found for $a/L\simeq 1$.
In fact, in this case the forces exerted by the constraints (maintaining the distance $a$ between the edges) are almost parallel to the graphene sheet, favoring the stretching emergence.

\begin{figure}[t]
\label{fig:bending flat1}
\centering
\includegraphics[width=.49\textwidth]{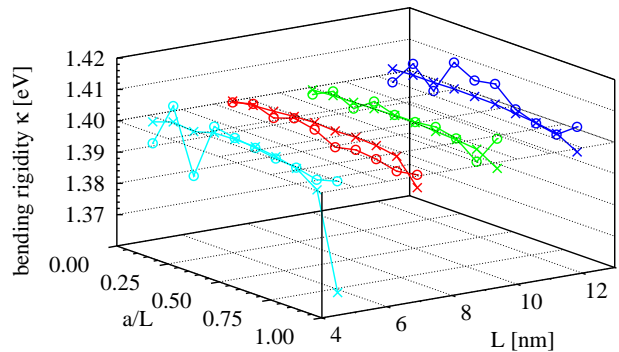}
\caption{\label{fig:bending flat}(Color online) Comparison between the bending rigidity $\kappa$ computed through ${U}_{b}= E_{o}^{bended}-E_{o}^{flat}$ (open circles) and ${U}_{b}= E_{relaxed}^{bended}-E_{straightened}^{flat}$ (crosses). The maximum deviation is less than the $1.5\%$.}
\end{figure}

This intriguing result opens the problem of how to disentangle bending and stretching features. As shown in Fig.\ref{fig:bending rigidity}, this is especially important in the limit of small deformations, a situation of considerable practical interest. To this aim we have defined a proof-of-concept computational procedure based on the virtual process of
straightening (or unbending) of a given relaxed and bended ribbon: atoms are projected from such a configuration onto a plane by conserving all the 1nn bond lengths and all the 2nn planar angles. The process recovers a planar configuration, still maintaining all the details about any possible stretching (in-plane strain field); the corresponding energy $E_{straightened}^{flat}$ is straightforwardly evaluated by means of TB-AS.
The bending rigidity $\kappa$ can be consequently determined by using  ${U}_{b}= E_{relaxed}^{bended}-E_{straightened}^{flat}$: the results are shown in Fig.\ref{fig:bending flat} (crosses) where we also report $\kappa$ as obtained by ${U}_{b}= E_{o}^{bended} -E_{o}^{flat}$ (open circles). The comparison points out  a good agreement between the two different approaches
since stretching features are either at all non considered (open circles) or included in both the bended and flat configurations (crosses) so as to compensate.
It is interesting to note that the constant trend of $\kappa$  versus $ a $ and $ L $ has been found 
similar to Fig.\ref{capt:72}.
In other words, we have proved that the evaluation of $ \kappa $ through the energy term ${U}_{b}= E_{relaxed}^{bended}-E_{relaxed}^{flat}$ is not correct since it is corrupted by a strain energy amount which is not directly related to the bending process. 
The energy  due to the sole stretching field (induced by the bending process) can be accordingly defined as $E_{straightened}^{flat}-E_{relaxed}^{flat}$. The demonstration that such an energetic contribution corresponds only to stretching relies on the fact that both the terms $ E_{straightened}^{flat} $ and $ E_{relaxed}^{flat} $ have been evaluated on flat ribbons through TB atomistic simulations.   

\begin{figure}[t]
\label{fig:train map}
\includegraphics[width=.48\textwidth]{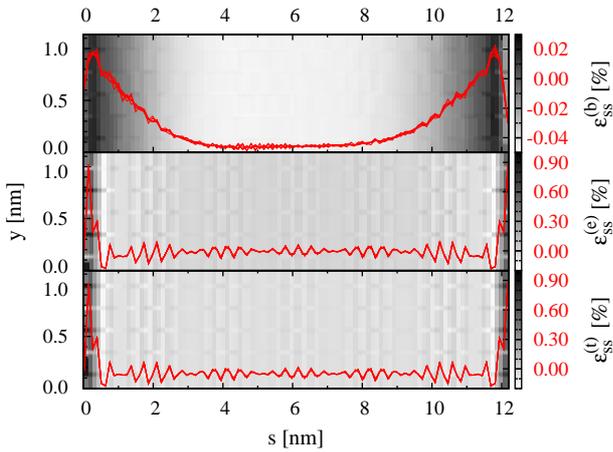}
\caption[.]{\label{fig:strain map} (Color online) $ \varepsilon_{ss}^{(b)} $ (strain induced by the bending), $ \varepsilon_{ss}^{(e)} $ (strain induced by the edges) and $ \varepsilon_{ss}^{(t)} $ (total strain) versus $ s $ (red curves) . The gray scale map in background represents the same quantities in the $sy$-space for $ L=12$ nm and $ a/L=0.95 $}
\end{figure}

A further evidence of the stretching emergence can be derived from Fig.\ref{fig:strain map} where the strain is calculated  along the arc of length $ L $ (corresponding to the dashed line in Fig.\ref{capt:geometri3D}), labeled by the coordinate $s$. We can calculate three strain fields $ \varepsilon_{ss}^{(b)} $, $ \varepsilon_{ss}^{(e)} $ and $ \varepsilon_{ss}^{(t)} $ which are respectively defined as the relative difference between: (i) the relaxed and straightened configuration (energy $E_{straightened}^{flat}$) and the flat relaxed configuration (energy $E_{relaxed}^{flat}$); (ii) the flat relaxed configuration (energy $E_{relaxed}^{flat}$) and the flat unrelaxed configuration (energy $E_{o}^{flat}$); (iii) the relaxed and  straightened configuration (energy $E_{straightened}^{flat}$) and the flat unrelaxed configuration (energy $E_{o}^{flat}$). 
While  the strain $ \varepsilon_{ss}^{(b)} $ is only due to bending, the term $ \varepsilon_{ss}^{(e)} $ is induced by the presence of the edges (finite nanoribbon) in a flat configuration. The quantity $ \varepsilon_{ss}^{(t)} $ represents the total strain induced by the relaxation of the bended ribbon with reference to the ideal graphene sheet. We observed with good accuracy the validity of the relation $ \varepsilon_{ss}^{(t)}=\varepsilon_{ss}^{(b)}+\varepsilon_{ss}^{(e)} $, further proving that the total strain in a bended ribbon is the sum of two different contributions: the first one ($ \varepsilon_{ss}^{(b)} $) is directly related to the bending process and the second one ($ \varepsilon_{ss}^{(e)} $) is originated by edges effects, i.e. by the finite size of the nanoribbon. Although the first term seems to be quite negligible with respect to the second one, the previous energetic analysis reveals that both contributions are essential in order to explain the discrepancies between continuum and atomistic results.

\section{Conclusions}
In conclusion, we offered robust arguments suggesting that the correct value for the bending rigidity of a carbon nanoribbon corresponds to $\kappa=~1.40$ eV, as calculated either through $  {U}_{b}=E_{o}^{bended} -E_{o}^{flat}$ or through ${U}_{b}=E_{relaxed}^{bended}-E_{straightened}^{flat}$. On the other hand, the relation ${U}_{b}= E_{relaxed}^{bended}-E_{relaxed}^{flat}$ leads to incorrect results because of the emergence of a stretching field $ \varepsilon_{ss}^{(t)} $. We have further proved that such an in-plane strain field can be decomposed in a first contribution $ \varepsilon_{ss}^{(b)} $ due to the actual bending and a second one $ \varepsilon_{ss}^{(e)} $ due to the edges effects.

SG and LC acknowledge financial support by COSMOLAB (Cagliari, Italy) and MATHMAT project (Universit\`a di Padova, Italy), respectively. We also acknowledge  COSMOLAB (Cagliari, Italy) and CASPUR (Rome, Italy) for computational support.

\bibliographystyle{apsrev}

\begin{thebibliography}{99}

\bibitem{novoselov}
A. K. Geim and K. S. Novoselov, Nature Materials \textbf{6} 183 (2007). 

\bibitem{RMP}
A. H. Castro Neto, F. Guinea, N. M. R. Peres, K. S. Novoselov and A. K. Geim , Rev. Mod. Phys. \textbf{81}, 109 (2009).

\bibitem{meyer}
J. C. Meyer, A. K. Geim, M. I. Katsnelson, K. S. Novoselov, T. J. Booth  and  S. Roth, Nature (London) \textbf{446}, 60 (2007).

\bibitem{ishi}
M. Ishigami, J. H. Chen, W. G. Cullen, M. S. Fuhrer and E. D. Williams, NanoLetters \textbf{7}, 1643 (2007).

\bibitem{res1}
J. S. Bunch,  A. M. van der Zande, S. S. Verbridge, I. W. Frank, D. M. Tanenbaum, J. M. Parpia, H. G. Craighead and P. L. McEuen, Science \textbf{315}, 490 (2007).

\bibitem{res2}
J. Atalaya, A. Isacsson and J. M. Kinaret, NanoLetters \textbf{8}, 4196 (2008).

\bibitem{schniepp}
H.C. Schniepp, K. N. Kudin, J.-L. Li, R. K. Prud'homme, R. Car, D. A. Saville and I. A. Aksay, ACS Nano {\bf 2}, 2577 (2008). 

\bibitem{kosy}
D. V. Kosynkin, A. L. Higginbotham, A. Sinitskii, J. R. Lomeda, A. Dimiev, B. K. Price  and  J. M. Tour, Nature (London) \textbf{458}, 872 (2009).

\bibitem{santos}
H. Santos, L. Chico, and L. Brey, Phys. Rev. Lett. \textbf{103}, 086801 (2009).

\bibitem{colombo}
L. Colombo, Riv. Nuovo Cimento \textbf{28}, 1 (2005).

\bibitem{xu} 
C.H. Xu, C. Z. Wang, C. T. Chan and K. M. Ho, J. Phys.: Condens. Matter {\bf 4}, 6047 (1992).

\bibitem{cadelano}
E. Cadelano, P. L. Palla, S. Giordano and L. Colombo, Phys. Rev. Lett. \textbf{102}, 235502 (2009).

\bibitem{good}
L. Goodwin, A. J. Skinner  and D. G. Pettifor, Europhys. Lett. \textbf{9}, 701 (1989).

\bibitem{green}
A. E. Green and W. Zerna, \textit{Theoretical Elasticity} (Oxford University Press, Oxford, 1954).

\bibitem{landau}
L.D. Landau and E.M. Lifschitz, {\em Theory of Elasticity} (Butterworth Heinemann, Oxford, 1986).

\bibitem{do carmo}
M. P. do Carmo, {\em Differential Geometry of Curves and Surfaces} (Prentice-Hall, New Jork, 1976)

\bibitem{arroyo3}
Q. Lu, M. Arroyo, and R. Huang, J. Phys. D: Appl. Phys. {\bf 42}, 102002 (2009).

\bibitem{grad} 
I. S. Gradshteyn and I. M. Ryzhik, {\it Table of integrals, series and products} (San Diego, Academic Press, 1965)

\bibitem{abra} 
M. Abramowitz and I. A. Stegun, \textit{Handbook of Mathematical Functions} (New York, Dover Publication, 1970)

\bibitem{tu}
Z. C. Tu, Z. C. Ou-Yang, J. Comput. Theor. Nanosci., \textbf{5} 422 (2008).

\bibitem{kudin}
K. N. Kudin, E. Scuseria and B. I. Yakobson, Phys. Rev. B {\bf 64}, 235406 (2001).

\bibitem{galli}
J.-Y. Raty, G. Galli, C. Bostedt, T. W. van Buuren, and L. J. Terminello, Phys. Rev. Lett. \textbf{90}, 037401(2003).

\bibitem{Bogana}
M.P. Bogana, and L.Colombo, Appl. Phys. A \textbf{86}, 275 (2007).
  	
\bibitem{Yamaguchi}
Y. Yamaguchi, L. Colombo, P. Piseri, L. Ravagnan, and P. Milani, Phys. Rev. B \textbf{76}, 134119 (2007).

\end{thebibliography}

\end{document}